\documentclass[12pt,a4paper]{article}

\usepackage{graphicx,epsfig,pstricks,rotating}
\usepackage{amssymb}
\usepackage{amsfonts}
\usepackage{amsmath}

\begin{document}
\textwidth=135mm
 \textheight=200mm
\begin{center}
{\bfseries Magnetic field buoyancy in accretion disks of young stars
\footnote{{\small Talk at the XX International Scientific Conference of Young Scientists and Specialists (AYSS-2016), JINR, Dubna, March 14 - 18,
2016.}}}
\vskip 5mm
S. A. Khaibrakhmanov$^{\dag,\ddag}$, A. E. Dudorov$^\ddag$
\vskip 5mm
{\small {\it $^\dag$  Kourovka astronomical observatory, Ural federal university, Ekaterinburg 620000, Russia}} \\
{\small {\it $^\ddag$ Theoretical physics department, Chelyabinsk state university, Chelyabinsk 454001, Russia}}
\\
\end{center}
\vskip 5mm
\centerline{\bf Abstract}
Buoyancy of the fossil magnetic field in the accretion disks of young stars is investigated. It is assumed that the Parker instability leads to the formation of slender flux tubes of toroidal magnetic field in the regions of effective magnetic field generation. Stationary solution of the induction equation is written in the form in which buoyancy is treated as the additional mechanism of the magnetic flux escape. We calculate the fossil magnetic field intensity in the accretion disks of young T~Tauri stars for the cases when radius of the magnetic flux tubes $a_{\mathrm{mft}} = 0.1H$, $0.5 H$ or $1H$, where $H$ is the accretion disk height scale. Calculations show that the buoyancy limits toroidal magnetic field growth, so that its strength is comparable with the vertical magnetic field strength for the case $a_{\mathrm{mft}}=0.1H$.
\vskip 10mm

 \section{Introduction}
Observations indicate that magnetic fields are ubiquitous in cosmos. Table~\ref{table1} shows typical values of the magnetic field strength in some cosmic objects. There are several methods for the determination of the magnetic field intensity~\cite{zeldovich_book}. The strength of the magnetic field component along the line of sight from the source to the observer can be determined using measurements of Zeeman splitting of spectral lines. Another method is measurement of Faraday rotation of the polarization plane of the radiation. Polarization measurements of synchrotron emission also give information about the magnetic field strength. The intensity of the protosolar nebula magnetic field is inferred from measurements of meteorites magnetization~\cite{levy78}.  The direction of the magnetic field vector in plane of the sky can be obtained from the analysis of the polarization of star light scattered by non-spherical dust grains aligned with the magnetic field. Polarization of thermal radiation of non-spherical dust grains also gives information about the magnetic field geometry.

   \begin{table}[h]
 	\caption{Typical magnetic field strengths in cosmic objects}
 	\label{table1}
 	\begin{center}
 		\begin{tabular}{|c|c|}
 			\hline
 			{\bf Object} & {\bf Magnetic field strength, G} \\
 			\hline
 			interstellar medium~\cite{spitzer_book} & $10^{-6}-10^{-5}$\\
 		    protosolar nebula~\cite{levy78} & $0.01-10$\\
 		    FU Orionis accretion disk~\cite{donati05} & $10^3$\\
 		    T~Tauri stars~\cite{krull07} & $1000-3000$ \\
 			white dwarfs~\cite{wd_mf} & $10^5-10^9$ \\
 			neutron stars~\cite{kaspi10} & $10^{12}-10^{14}$\\
 			\hline
 		\end{tabular}
 	\end{center}
 \end{table} 

Seed magnetic field in cosmos is generated by charge separation in plasma where density and pressure gradients are not collinear. This is so-called Biermann battery mechanism~\cite{biermann}. Hydromagnetic dynamo enhances seed magnetic field in process of turbulent cyclonic motions and differential rotation in conducting plasma~\cite{parker55}. 

Another theory of stellar magnetic field origin is the theory of fossil magnetic field~\cite{dud95, fmft}. It is based on the observations and numerical simulations of star formation in magnetic rotating cores of molecular clouds (protostellar clouds). Stars form in process of gravitational instability and subsequent collapse of the protostellar clouds under conditions of partial ionization. The centrifugal and electromagnetic forces lead to the formation of a disk around the protostar during the cloud collapse. The magnetic flux of protostellar clouds is conserved partially during the star formation. This means that young stars with accretion disks form with the fossil magnetic field, i.e. their magnetic field is the remnant of the magnetic field of the protostellar clouds.

The accretion disks of young stars are disks of gas and dust~\cite{williams11}. They are geometrically thin and optically thick. The disks have typical masses $0.001-0.1M_{\odot}$ and typical radii 100-1000~AU\footnote{$1\,\mathrm{AU} = 1.5\times 10^{13}$~cm}. Matter accretes from the disk onto the star with mass rates $\dot{M}=(10^{-9}-10^{-6})M_{\odot}\,\mathrm{yr}^{-1}$ within $10^6-10^7$~yrs. Donati et al \cite{donati05} detected magnetic field strength $\sim$1000 Gs in the accretion disk of FU Orionis system using measurements of Zeeman splitting of spectral lines. Measurements of the polarized 1.25~mm continuum emission from HL~Tau disk have shown that the magnetic field in the disk has both toroidal and poloidal components~\cite{stephens14}.
  
In paper~\cite{fmfadys} (DK14, hereafter), we found  that extremely strong azimuthal magnetic field is generated in the inner regions of the accretion disks of T~Tauri stars, where thermal ionization operates. Therefore, the problem is what process limits further generation of the azimuthal magnetic field. Strong regular azimuthal magnetic field splits into separated toroidal magnetic flux tubes (MFTs) due to the Parker instability~\cite{parker_book}. Influence of the buoyancy of MFTs on magnetic field strength in accretion disks has been investigated by \cite{campbell_book, zhilkin12}.

We calculate the geometry and intensity of the fossil magnetic field in accretion disks of T~Tauri stars taking into account the toroidal magnetic field escape in the process of buoyancy. In section \ref{sec:model}, we describe a model of the accretion disk. Calculations results are presented in section \ref{sec:results}. We summarize and discuss our results in section \ref{sec:summ}.

 \section{Model}
 \label{sec:model}
Let us describe briefly the basic model of the accretion disk. Details can be found in DK14. We consider geometrically thin and optically thick accretion disk of young star. Cylindrical coordinate system $\left\{r,\, \varphi,\, z\right\}$ is used. Disk rotation axis is directed along $z$-axis. Initial magnetic field is poloidal. Gravity of the disk is neglected comparing to the gravity of the star.
 
The radial structure of the disk is described by equations of Shakura and Sunyaev~\cite{ss73}. It is considered that angular momentum is transported radially by turbulent stresses. Turbulent viscosity coefficient is written as $\nu_t = \alpha V_s H$, where $\alpha<1$ is the non-dimensional parameter characterizing turbulence efficiency, $V_s$ is the sound speed, $H$ is the accretion disc height scale. Gas temperature is determined from the balance between heating by turbulent stresses and radiative cooling. The disk is in the hydrostatic equilibrium, $V_z=0$.
 
Ionization fraction is calculated from Spitzer's equation~\cite{spitzer_book} of collisional ionization taking into account cosmic rays, X-rays and decay of radioactive elements. Radiative recombinations and recombinations on the dust grains are considered. In the regions where temperature $T>400$~K, we calculate thermal ionization of hydrogen and metals using the Saha equation.
 
We write the induction equation in the following form~\cite{dud89}
 \begin{equation}
 \frac{\partial {\bf B}}{\partial t} = \mbox{ rot}\left[{\bf V} + {\bf V}_{\rm{mad}},\, {\bf B}\right]  + \rm{rot}\left(\eta_{\rm{od}}\mbox{rot}\,{\bf B}\right) +  \rm{rot}\left[{\bf V}_B,\,{\bf B}_t\right],\label{Eq:Ind}
 \end{equation}
 where ${\bf B}=\left(B_r,\, B_{\varphi},\, B_z\right)$ is the magnetic field vector, ${\bf V}=\left(V_r,\, V_{\varphi},\, 0\right)$ the gas velocity vector, the ${\bf V}_{\rm{mad}}$ the stationary velocity of magnetic ambipolar diffusion (MAD), $\eta_{\rm{od}}$ the ohmic diffusivity,  ${\bf V}_B = \left(0,\, 0,\, V_B\right)$ the rise velocity of the toroidal magnetic flux tubes, ${\bf B}_t = \left(0, \, B_{\varphi}, \, 0\right)$ the toroidal magnetic field.
 
Velocity ${\bf V}_{\rm{mad}}$ is the velocity of the plasma drift through the neutral gas under the action of electromagnetic force
 \begin{equation}
 {\bf V}_{\rm{mad}} = \frac{\left[\mbox{rot}\,{\bf B}, {\bf B}\right]}{4\pi  \mu_{\rm{in}}n_{\rm{i}}n_{\rm{n}}\langle\sigma v\rangle_{\rm{in}}}, \label{Eq:MADvelocity}
 \end{equation}
where $\mu_{\rm{in}}$ is the reduced mass for ion and neutral particles, $n_{\rm{i}}$ the ions number density, $n_{\rm{n}}$ the neutral particles number density, $\langle\sigma v\rangle_{\rm{in}}= 2\times 10^{-9}\,\rm{cm}^3\,\rm{s}^{-1}$ \cite{spitzer_book} the coefficient of momentum transfer in collisions between ions and neutrals. The Ohmic diffusion (OD) of the magnetic field is due to currents dissipation in plasma with finite conductivity
\begin{equation}
	\eta_{\rm{od}} = \frac{c^2}{4\pi\sigma},\label{Eq:nu_m}
\end{equation}
$c$ is the speed of light, $\sigma$ the electrical conductivity. In the model approximations, $V_{\varphi}$ is determined from the centrifugal balance,
\begin{equation}
 V_{\varphi} = \sqrt{\frac{GM}{r}}\left(1 + \frac{z^2}{r^2}\right)^{-3/4},
\end{equation} 
where $G$ is the gravitational constant, $M$ the mass of the star. Dependence  $V_r(z)$ is not determined in the model, and derivative $\partial V_r/\partial z$ is estimated as $V_r/z$. Stationary solution of the equation~(\ref{Eq:Ind}) is
 \begin{eqnarray}
 	B_z &=& B_{z0}\frac{\Sigma}{\Sigma_0}, \label{Eq:BzSol}\\
 	B_r &=& -\frac{V_rzB_z}{\eta_{\mathrm{mad}} + \eta_{\mathrm{od}}},\label{Eq:BrSol}\\
 	B_{\varphi} &=& -\frac{3}{2}\left(\frac{z}{r}\right)^2\frac{V_{\varphi}z}{\eta_{\mathrm{tot}}}B_z + \frac{1}{2}\left(\frac{z}{r}\right)\frac{V_{\varphi}z}{\eta_{\mathrm{tot}}}B_r,\label{Eq:BphiSol}
 \end{eqnarray}
 where $\Sigma=2\int_{0}^H\rho dz$ is the surface density, $\rho$ the density, $B_{z0}$ and $\Sigma_0$ are values of $B_z$ and $\Sigma$ at the outer edge of the disk,
 \begin{equation}
 	\eta_{\mathrm{tot}} = \eta_{\mathrm{mad}} + \eta_{\mathrm{od}} + |V_B|H\label{Eq:eta_tot}
 \end{equation}
 is the ``effective'' diffusivity describing contribution of MAD, OD and buoyancy.  MAD diffusivity
 \begin{equation}
 \eta_{\rm{mad}} = \frac{B^2}{4\pi  \mu_{\rm{in}}n_{\rm{i}}n_{\rm{n}}\langle\sigma v\rangle_{\rm{in}}},\label{Eq:eta_mad}
 \end{equation}

Expression~(\ref{Eq:eta_tot}) shows that the buoyancy can be considered as the additional mechanism of the magnetic field diffusion.
 
Velocity $V_B$ is determined from the balance between buoyancy and drag forces. In the aerodynamic case, stationary MFT rise velocity~\cite{parker_book}
\begin{equation}
	V_B = V_a\left(\frac{\pi}{C_D}\right)^{1/2}\left(\frac{a_{\rm{mft}}}{H}\right)^{1/2}\left(\frac{z}{H}\right)^{1/2},\label{Eq:V_b_a}
\end{equation}
where
\begin{equation}
	V_a = \frac{B}{\sqrt{4\pi\rho}}
\end{equation}
is the Alfv{\'e}n velocity, $C_D\simeq 1$ the drag coefficient, $a_{\mathrm{mft}}$ the MFT cross-section radius. In the turbulent drag case~\cite{pneuman72}, 
\begin{equation}
	V_B = V_a \left(\frac{\beta}{2}\right)^{-1/2}\alpha^{-1/3}\left(\frac{a_{\rm{mft}}}{H}\right)\left(\frac{z}{H}\right)^{2/3},\label{Eq:Vb_t}
\end{equation}
where $\beta = 8\pi\rho V_s^2/B^2$ is the plasma parameter.

\section{Results} 
 \label{sec:results}
In this section, we present results of calculations of accretion disk structure and its fossil magnetic field for solar mass T~Tauri star, $M=1\,M_{\odot}$, $\dot{M}=10^{-7}\,M_{\odot}\,\mathrm{yr}^{-1}$, $\alpha=0.01$. In the calculations, cosmic rays ionization rate $\xi_0=10^{-17}\,\rm{s}^{-1}$ and attenuation length $R_{\rm{CR}}=100\,\rm{g}\,\rm{cm}^{-2}$, stellar X-ray luminosity $L_{\rm{XR}}=10^{30}\,\rm{erg}\,\rm{s}^{-1}$, mean dust particle radius $a_{\rm{d}}=0.1\,\mu\rm{m}$.

\begin{figure}[t]
	\centerline{
		\includegraphics[width = 12.5cm]{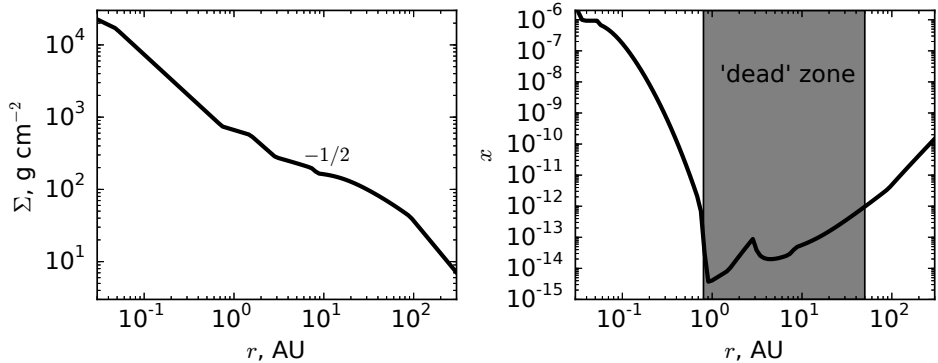}}
	\caption{Left: radial profile of the surface density. Right: radial profile of the ionization fraction in the midplane of the disk.}
	\label{fig1}
\end{figure}

In Fig.~\ref{fig1}, we plot surface density (left panel) and midplane ionization fraction (right panel) versus radial distance in the disk. The surface density decreases with distance as $\Sigma \propto r^{-1/2}$ at $1\,\mathrm{AU}<r<100$~AU. Ionization fraction $x>10^{-12}$ due to the thermal ionization at $r<0.8$~AU. Thermal ionization is not efficient at $r>0.8$~AU and ionization fraction decreases with distance. Efficient cosmic rays ionization increases $x$ at $r>50$~AU. Gray region in the right panel is the so-called ``dead'' zone~\cite{gammie96}, i.e. the region of very low ionization fraction, $x<10^{-12}$, where magnetic diffusion limits the generation of magnetic field. 

\begin{figure}[t]
	\centerline{
		\includegraphics[width = 12.5cm]{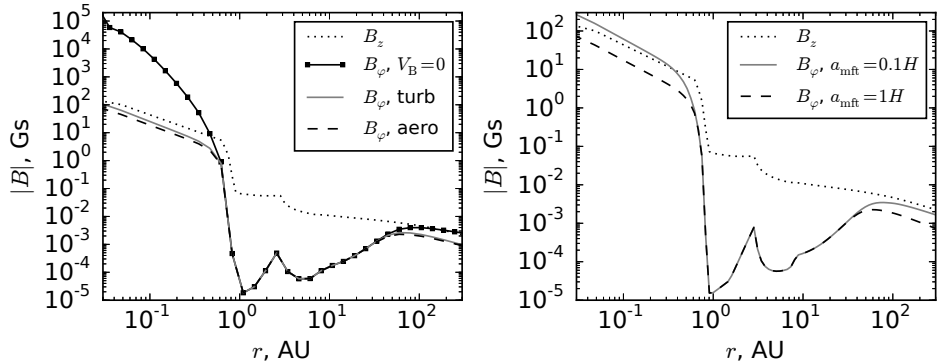}}
	\caption{Radial profiles of the vertical and azimuthal magnetic field components at $z=0.5H$. Left: dependence on drag type, right: dependence on MFT radius.}
	\label{fig2}
\end{figure}

In Fig.~\ref{fig2}, we show radial profiles of $B_z$ (dotted line) and $B_{\varphi}$ calculated with and without buoyancy. In the left panel, we investigate dependence on drag type (gray solid line~--~turbulent, black dashed line~--~aerodynamic) for given MFT radius $a_{\mathrm{mft}}0.5H$. Black line with squares depicts $B_{\varphi}$ calculated with OD and MAD only. In the right panel, dependence on MFT radius is examined for the case of turbulent drag (gray line: $a_{\mathrm{mft}}=0.1H$, black dashed line: $a_{\mathrm{mft}}=1H$). 

Figure~\ref{fig2} shows that $B_{\varphi}$ is nearly two times smaller than $B_z$ at $r<0.8$~AU and $r>50$~AU due to the buoyancy in the case $a_{\mathrm{mft}}=0.5H$. Otherwise, $B_{\varphi}$ is enormously large in these regions. Contributions of aerodynamic and turbulent drags are comparable. This is because the Reynolds number is of order of unity in considered case.
 Inside the ``dead'' zone, $0.8\,\mathrm{AU}<r<50$~AU, $B_{\varphi}\ll B_z$ because of OD and MAD. The right panel shows that the more MFT radius the less $B_{\varphi}$. In the case $a_{\mathrm{mft}}=0.1H$, the toroidal magnetic field strength is comparable with the vertical magnetic field strength. Maximal MFT speed in this case is of order of the Alfv{\'e}n speed that is equal to $0.21\,\rm{km}\,\rm{s}^{-1}$ at $r=0.45$~AU.
 
\section{Summary and discussion}
\label{sec:summ}
We investigated the buoyancy of the fossil magnetic field in the accretion disks of young T~Tauri stars.  It is considered that the toroidal magnetic field splits into the slender magnetic flux tubes that rise to the surface of the disk. Stationary velocity of the MFT is determined from the balance between buoyancy force and drag force. Turbulent and aerodynamic drags have been examined. Calculations were carried out for the cases when radius of the MFTs is $0.1H$, $0.5H$ or $1H$.  

We show that the buoyancy can be treated as the additional mechanism of the magnetic flux diffusion. The buoyancy limits the toroidal magnetic field, so that its strength is comparable with the vertical magnetic field strength for the case when $a_{\mathrm{mft}}=0.1H$. Contributions of the turbulent and aerodynamic drags are comparable for the considered parameters.

We propose that rising MFTs can appear as the periodic outflows from the accretion disks. Velocity of such outflows is comparable with local Alfv{\'e}n speed, $\sim (0.2-0.3)\,\rm{km}\,\rm{s}^{-1}$ for the considered parameters. In our upcoming paper, we aim to investigate the dynamics of the slender flux tubes in the accretion disks of young stars.

\section*{Acknowledgments}
S. Khaibrakhmanov is supported by Russian science foundation (project 15-12-10017). A. Dudorov is supported by Foundation of perspective scientific research of the Chelyabinsk state university (project 5/16).

\end{document}